\documentclass[11pt,a4paper]{article}

\usepackage[margin=1in]{geometry}
\usepackage[T1]{fontenc}
\usepackage[utf8]{inputenc}
\usepackage{lmodern}
\usepackage{setspace}
\usepackage{graphicx}
\usepackage{microtype}
\usepackage[hidelinks]{hyperref}
\usepackage{tabularx}
\usepackage{booktabs}

\graphicspath{{Figures/}}

\newcommand{\MainHeading}[1]{%
  \section*{#1}%
}
\newcommand{\SubHeading}[1]{%
  \subsection*{#1}%
}
\newcommand{\LiteralHeading}[1]{%
  \subsubsection*{#1}%
}
\newcommand{\code}[1]{\texttt{\detokenize{#1}}}
\newcommand{\refnum}[1]{\hyperlink{ref:#1}{#1}}
\newcommand{\refone}[1]{[\refnum{#1}]}
\newcommand{\reftwo}[2]{[\refnum{#1},\refnum{#2}]}
\newcommand{\refthree}[3]{[\refnum{#1},\refnum{#2},\refnum{#3}]}
\newcommand{\refrange}[2]{[\refnum{#1}--\refnum{#2}]}

\begin{document}

\begin{center}
{\LARGE\bfseries Research trends in music-based interventions in neonatal intensive care units: a text mining and topic modeling study\par}
\vspace{1em}
{\large Min young Choun\textsuperscript{1}, Mijeong Kim\textsuperscript{2}, Soo Ji Kim\textsuperscript{1,*}\par}
\vspace{0.5em}
{\normalsize
\textsuperscript{1}Graduate Program in Music Therapy, Ewha Womans University, Seoul, Republic of Korea\par
\textsuperscript{2}Department of Statistics, Ewha Womans University, Seoul, Republic of Korea\par
\textsuperscript{*}Correspondence: Soo Ji Kim, specare@ewha.ac.kr; ORCID: 0000-0002-7279-1804\par
}
\end{center}

\MainHeading{Abstract}

\SubHeading{Background}

Music-based interventions are increasingly used in neonatal intensive care units (NICUs) to support the physiological stability, developmental outcomes, and emotional well-being of preterm infants and their families. However, the literature remains heterogeneous in terms of intervention type, provider role, and research focus. This study aimed to examine research trends and thematic directions in NICU music-based intervention studies using text mining techniques.

\SubHeading{Methods}

A total of 83 abstracts from peer-reviewed studies published between 1998 and 2025 were analyzed. Text mining procedures included text preprocessing, RAKE-based keyphrase extraction, keyword frequency analysis, temporal trend analysis, comparison by intervention type, and latent Dirichlet allocation (LDA) topic modeling. To determine the optimal number of topics, the \code{FindTopicsNumber} function from the \code{ldatuning} package was applied using the \code{CaoJuan2009}, \code{Arun2010}, and \code{Deveaud2014} metrics.

\SubHeading{Results}

The number of NICU music-based intervention studies increased steadily over time, with a marked acceleration in recent years; nearly half of the included studies (38/83) were published from 2020 onward. Early studies primarily focused on passive music listening and short-term physiological outcomes, whereas more recent studies increasingly examined active and interactive approaches, such as singing, live music, and parent-involved interventions. Keyword analysis showed that the literature initially emphasized physiological stability and behavioral responses in preterm infants, but later expanded to include neurodevelopmental outcomes, parental emotional well-being, and parent-infant interaction. Comparative analysis further indicated that music medicine (MM) studies mainly focused on passive auditory stimulation and immediate physiological outcomes, whereas music therapy (MT) studies addressed a broader range of developmental, relational, and psychosocial topics. Topic modeling identified four major themes, with parent-involved physiological regulation and stress reduction emerging as the most frequent dominant topic (28/83 studies), followed by growth and developmental outcomes of music interventions, immediate physiological responses to music stimuli, and the developmental and psychosocial effects of music therapy.

\SubHeading{Conclusions}

Research on NICU music-based interventions is evolving toward a more multidimensional and interdisciplinary direction. The field has expanded from a primary focus on immediate physiological stabilization to broader developmental, relational, and psychosocial goals. For future clinical application, it will be important to strengthen the role of certified music therapists, clarify the conceptual distinction between music therapy and music medicine, and promote interdisciplinary collaboration in NICU care.

\SubHeading{Keywords}

neonatal intensive care unit; music-based intervention; music therapy; music medicine; text mining; topic modeling

\MainHeading{Background}

Music-based interventions have become increasingly recognized as evidence-informed complementary approaches in neonatal intensive care units (NICUs). Previous studies have suggested that music may help stabilize physiological functioning, improve sleep, reduce pain-related distress, and support feeding and developmental regulation in preterm infants \refrange{1}{3}. As concerns continue regarding the limitations and adverse effects of pharmacological treatment, interest in non-pharmacological strategies has expanded in neonatal care. Among sensory-based interventions used in NICUs, music is particularly notable because it is non-invasive, emotionally meaningful, and adaptable to both infant and family needs \reftwo{4}{5}.

Preterm infants are especially vulnerable to repeated medical procedures, environmental noise, and sensory disruption during a critical period of neurological development. These stressors may adversely affect autonomic regulation, auditory maturation, and later developmental outcomes \refrange{6}{8}. Parents of preterm infants likewise experience substantial psychological burden, including anxiety, depression, and emotional isolation, which may interfere with bonding and caregiving \reftwo{9}{10}. In this context, music-based interventions are clinically meaningful because they may support not only infant physiological regulation but also parental emotional well-being and parent-infant interaction. Reported benefits include improved cardiorespiratory patterns, sleep, feeding behavior, and bonding, as well as reduced anxiety and shorter hospital stays \refrange{11}{13}.

Despite these promising findings, the NICU music intervention literature remains heterogeneous. Studies differ in intervention type, delivery mode, provider role, and outcome focus. In particular, music therapy (MT), which is generally delivered by credentialed music therapists, and music medicine (MM), which often involves recorded or externally delivered music stimulation, are frequently discussed within the same body of literature, even though they differ conceptually and clinically \refrange{14}{16}. Previous reviews have reported beneficial effects of music interventions in NICU settings, but many have focused primarily on specific outcomes or limited intervention types \refthree{13}{17}{18}. As a result, the broader thematic structure of the field, including temporal shifts and differences between MT and MM, remains insufficiently clarified.

Text mining offers a useful approach to addressing these limitations by enabling large-scale, data-driven analysis of published literature while reducing subjective bias. In particular, keyphrase-based analysis can capture semantically meaningful multi-word concepts more effectively than single-word approaches, and topic modeling can reveal latent thematic structures across studies \reftwo{19}{20}. These methods are well suited to identifying major research domains, tracing changes in research focus over time, and comparing conceptual emphases across intervention types. Given the diversity of NICU music intervention research, a text mining approach may provide a more comprehensive understanding of how the field has developed across physiological, developmental, and psychosocial domains.

Accordingly, the present study applied text mining techniques to abstracts of published NICU music-based intervention studies from 1998 to 2025. Using RAKE-based keyphrase extraction, temporal keyword analysis, intervention-type comparison, and latent Dirichlet allocation topic modeling, this study aimed to map research trends and identify the major thematic structures of the field. In particular, this study sought to clarify how research topics have shifted over time and how emphases differ between MT- and MM-related studies.

This study addressed the following research questions:

\begin{enumerate}
\setlength{\itemsep}{4pt}
  \item What are the trends in research on music-based interventions in NICUs?
  \item What are the main research topics in NICU music-based intervention studies?
\end{enumerate}

\MainHeading{Methods}

\SubHeading{Study design}

This study used text mining and topic modeling to analyze research trends and thematic structures in published NICU music-based intervention studies.

\SubHeading{Data selection}

Literature was retrieved from Google Scholar, PubMed, and EBSCO using combinations of terms related to the NICU context, including infant intensive care, NICU, neonatal care, preterm, and premature infant care, together with music-related terms such as music intervention, music therapy, music, singing, listening, and playing. The search covered publications from 1998 to February 2025.

Studies were included if they implemented music-based interventions with neonates in NICU settings. Both quantitative and qualitative studies were considered. Only peer-reviewed journal articles were included. Dissertations, non-English or non-Korean papers, and studies judged irrelevant on the basis of titles or abstracts were excluded. Studies were also excluded if they did not provide a sufficiently clear description of the music intervention. After screening and eligibility assessment, 83 studies were included in the final review. This final corpus was intentionally restricted to peer-reviewed NICU music-intervention studies so that the analysis would reflect a focused and clinically relevant body of literature rather than a broader but more heterogeneous set of records.

\SubHeading{Data analysis}

All analyses were conducted in R \refone{25}. The analytic workflow combined established text-mining algorithms with their R implementations, including UDPipe-based linguistic annotation, RAKE-based keyphrase extraction, tidytext-based tokenization and stopword filtering, LDA estimation, and topic-number tuning. The main R packages, key functions, and analytic roles are summarized in Table 1. Detailed preprocessing and keyword-filtering rules are provided in Additional file 1: Supplementary Method S1.

\begin{table}[ht]
\centering
\caption{Main R packages and functions used in the analytic workflow}
\label{tab:rpackages}
\small
\renewcommand{\arraystretch}{1.15}
\begin{tabularx}{\textwidth}{@{}p{2.8cm}p{2.8cm}p{3.4cm}X@{}}
\toprule
\textbf{Analytic step} & \textbf{Package} & \textbf{Key function} & \textbf{Role} \\
\midrule
Linguistic annotation and part-of-speech tagging & \code{udpipe} \refone{26} & \code{udpipe_annotate} & Performed token-level annotation and part-of-speech tagging based on the UDPipe pipeline. \\
Keyphrase extraction & \code{udpipe} \refone{26} & \code{keywords_rake} & Extracted RAKE-based multi-word keyphrases from the annotated abstracts. \\
Tokenization and stopword filtering & \code{tidytext} \refone{27} & \code{unnest_tokens} & Tokenized cleaned text and supported stopword removal before document-term matrix construction. \\
Topic modeling & \code{topicmodels} \refone{28} & \code{LDA} & Fitted the latent Dirichlet allocation model and estimated topic-term and document-topic distributions. \\
Topic-number tuning & \code{ldatuning} \refone{29} & \code{FindTopicsNumber} & Evaluated candidate topic numbers using the \code{CaoJuan2009}, \code{Arun2010}, and \code{Deveaud2014} metrics. \\
\bottomrule
\end{tabularx}
\normalsize
\end{table}

RAKE (Rapid Automatic Keyword Extraction) was used to extract keyphrases from the study abstracts. RAKE is a domain-independent and unsupervised keyword extraction method designed to identify keywords as sequences of one or more words from individual documents without requiring training data \refone{21}. This characteristic is important for the present study because clinically meaningful concepts in NICU music-intervention research are often expressed as multi-word units rather than as isolated single tokens. In contrast to simple single-word frequency analysis, which counts individual terms separately and may fragment conceptually unified expressions, RAKE first generates candidate phrases by splitting text at stopwords and punctuation and then weights each word according to its frequency and co-occurrence degree within those phrases. Candidate phrase scores are calculated from the summed scores of their constituent words, allowing recurrent and semantically cohesive phrases to be prioritized over merely frequent individual words \refone{21}.

\SubHeading{Keyword analysis procedure}

To identify major research topics and core keywords in NICU music intervention studies, the selected abstracts were used as the unit of analysis. During preprocessing, abbreviations and acronyms were expanded, and synonym normalization was performed to standardize semantically equivalent terms. For example, \code{neonate}, \code{newborn}, and \code{infants} were unified as \code{infant}, whereas \code{preterm} was standardized as \code{premature}. Expressions related to oxygen saturation, such as \code{SpO2} and \code{rSO2}, were normalized as \code{oxygen saturation}.

Part-of-speech tagging was conducted using the \code{udpipe} package, which implements the UDPipe annotation pipeline, and only nouns and adjectives identified by \code{udpipe_annotate} were retained as keyword candidates in order to focus on semantically meaningful content. RAKE was then applied with the \code{keywords_rake} function to extract keyphrases, allowing n-grams of up to four words. To improve keyword quality, custom stopwords were used to remove generic research terms and low-information words. In addition, single-word keywords were excluded so that only multi-word keyphrases were retained for analysis.

Nested phrases representing the same core concept were normalized into shorter, more general forms. For example, \code{stable premature infant} was reduced to \code{premature infant}, and \code{creative music therapy} and \code{live music therapy} were normalized as \code{music therapy}. This phrase-based approach was especially useful because many clinically meaningful expressions in this literature, such as \code{heart rate}, \code{kangaroo care}, and \code{mother anxiety}, are more interpretable as multi-word units than as separate single tokens. In addition, possessive markers such as apostrophe-\code{s} endings were removed during preprocessing to reduce superficial variation across equivalent expressions. As a result, a small number of extracted two-word labels may appear lexically simplified or less natural as surface English, even though the underlying concept remains unchanged. These should therefore be interpreted as normalized analytical keyphrases rather than as intended grammatical phrases from the original abstracts. For each extracted keyphrase, the number of documents in which it appeared (\code{n_doc}), total frequency, and mean RAKE score were calculated. These indicators were used to identify frequently occurring keyphrases and characterize the main research themes. Representative period-specific keywords are additionally summarized in Additional file 1: Table S1.

\SubHeading{Topic modeling procedure}

LDA-based topic modeling was conducted to identify latent thematic structures in the literature. Keywords extracted using RAKE were filtered to include only those with a document frequency of two or higher. Multi-word keyphrases were preserved as single semantic units by replacing spaces with underscores, and corresponding phrases in the abstracts were replaced with single tokens. Tokenization was performed using the \code{tidytext} package, stopwords were removed, and only tokens corresponding to the predefined keyword set were retained. A document-term matrix was then constructed based on keyword frequencies within each document.

To determine the optimal number of topics (\code{k}), the \code{FindTopicsNumber} function from the \code{ldatuning} package was applied using the \code{CaoJuan2009}, \code{Arun2010}, and \code{Deveaud2014} metrics. The \code{CaoJuan2009} metric evaluates topic density by examining similarity among topics and generally favors lower values when topics are better separated. The \code{Arun2010} metric compares topic-word structure with document-topic structure and also favors lower values when the latent structure is more stable. In contrast, the \code{Deveaud2014} metric is based on pairwise Jensen-Shannon divergence and favors higher values when topics are more distinct from one another \refrange{22}{24}. Because the corpus was relatively small and abstract-based, topic selection emphasized parsimony as well as statistical fit. Models with fewer topics merged clinically distinct domains, whereas models with more topics produced fragmented themes with limited interpretability and unstable assignment patterns across publication periods and intervention types. Considering the balance among these indices together with topic interpretability, the final number of topics was set to four. LDA modeling was then performed using the \code{topicmodels} package, and the probability distribution of terms within each topic (\code{beta} values) was estimated. In probabilistic terms, the model assumes that the probability of observing word \(w\) in document \(d\) is represented as a weighted mixture of topic-specific word distributions:

\[
P(w \mid d) = \sum_{t=1}^{K} P(w \mid t) P(t \mid d)
\]

where \(P(w \mid t)\) denotes the probability of word \(w\) under topic \(t\), and \(P(t \mid d)\) denotes the topic proportion for document \(d\). In the fitted LDA model, these document-topic proportions were represented by \code{gamma} values, and each abstract was assigned to the topic with the highest \code{gamma} value for the temporal topic-trend analysis. The top keywords for each topic were extracted to describe topic characteristics. Topic similarity was further examined by calculating cosine similarity based on keyword distributions across topics and visualized using a heatmap (Additional file 1: Figure S1). An additional exploratory analysis of effect-focused intervention phrases is described in Additional file 1: Supplementary Method S2.

\MainHeading{Results}

\SubHeading{Keyword analysis}

\LiteralHeading{Keyword frequency analysis}

Keyphrases were extracted from the abstracts using RAKE, and keyword frequency analysis was conducted based on the number of documents in which each keyword appeared (\code{n_doc}) and the total frequency of occurrence (\code{total_freq}). The most frequent RAKE-derived keyphrases are presented in Figure 1, and the top-ranked keyphrases with their summary statistics are shown in Table 2. Figure 1 reflects the total frequency of normalized multi-word keyphrases extracted from the abstracts rather than the number of included studies. Because the plotted labels reflect post-preprocessing keyphrases, a few may appear slightly abbreviated or less grammatically natural after possessive removal and phrase normalization.

Overall, the most prominent keyphrases indicate that the literature has concentrated on four broad domains: physiological regulation in preterm infants, developmental outcomes, parent-infant interaction, and behavioral or neurodevelopment-related responses. In particular, high-ranking phrases related to physiological stability, growth, and parental involvement suggest that NICU music intervention research has moved beyond simple exposure studies toward a broader developmental and relational focus (Table 2; Figure 1).

\begin{table}[ht]
\centering
\caption{Top 10 RAKE-derived keyphrases in the analyzed abstracts}
\label{tab:top10keywords}
\begin{tabularx}{\textwidth}{>{\raggedright\arraybackslash}X c c c}
\toprule
Keyphrase & \code{n\_doc} & \code{total\_freq} & Mean RAKE score \\
\midrule
\code{premature infant} & 50 & 192 & 2.40 \\
\code{music therapy} & 32 & 192 & 3.62 \\
\code{heart rate} & 15 & 43 & 2.41 \\
\code{respiratory rate} & 10 & 26 & 2.13 \\
\code{physiological parameters} & 7 & 23 & 2.19 \\
\code{weight gain} & 7 & 20 & 2.43 \\
\code{kangaroo care} & 6 & 36 & 2.52 \\
\code{behavioral state} & 4 & 10 & 2.26 \\
\code{mother anxiety} & 4 & 12 & 1.76 \\
\code{music intervention} & 4 & 12 & 2.19 \\
\bottomrule
\end{tabularx}
\end{table}

Overall, the keyword analysis indicates that NICU music-based intervention research has developed around four broad domains: physiological responses in preterm infants, parent-infant interaction, behavioral outcomes, and neurodevelopment.

\begin{figure}[ht]
\centering
\includegraphics[width=0.9\textwidth]{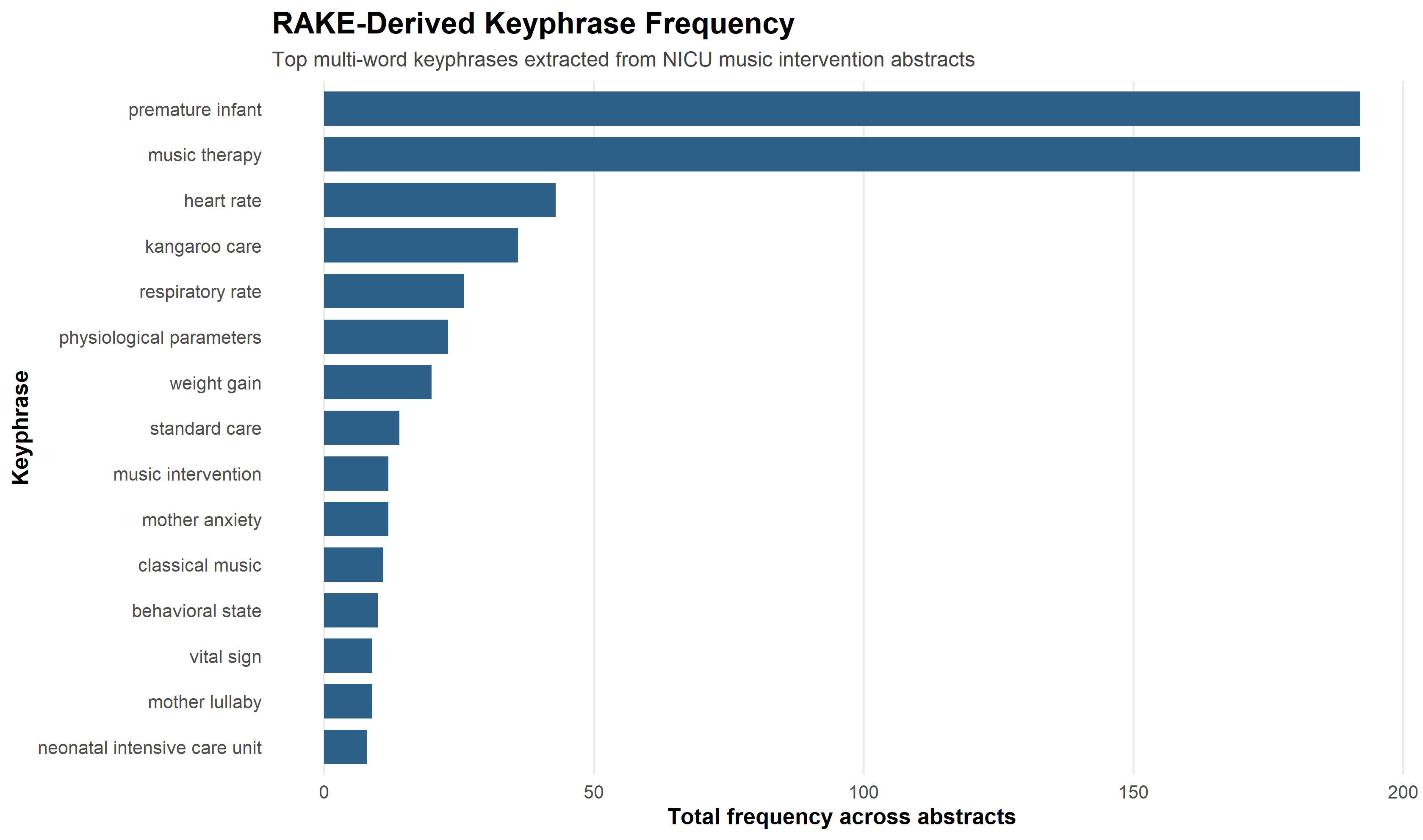}
\caption{RAKE-derived keyphrase frequencies from the analyzed abstracts. Bars indicate the total frequency of normalized multi-word keyphrases extracted from the abstracts, not the number of included studies. Labels should be interpreted as analytical phrase units after preprocessing and normalization rather than as full grammatical phrases from the original abstracts.}
\end{figure}

\LiteralHeading{Keyword frequency analysis by period}

To examine temporal changes in research topics, the included studies were divided into four publication periods based on publication year so that each period contained a roughly comparable number of studies, and keywords with a frequency of two or more were extracted for each period. Representative high-frequency keyphrases for each period are summarized in Table 3.

\begin{table}[ht]
\centering
\caption{Representative high-frequency keyphrases by publication period}
\label{tab:periodkeywords}
\begin{tabularx}{\textwidth}{l >{\raggedright\arraybackslash}X}
\toprule
Period & Representative high-frequency keyphrases \\
\midrule
Period 1 (1998--2015) & \code{premature infant}; \code{music therapy}; \code{heart rate}; \code{respiratory rate}; \code{kangaroo care}; \code{salivary cortisol}; \code{stress level} \\
Period 2 (2016--2020) & \code{premature infant}; \code{music therapy}; \code{heart rate}; \code{respiratory rate}; \code{mother lullaby}; \code{lullaby stage}; \code{music intervention} \\
Period 3 (2020--2022) & \code{music therapy}; \code{premature infant}; \code{heart rate}; \code{physiological parameters}; \code{mother anxiety}; \code{depressive symptoms}; \code{weight gain} \\
Period 4 (2022--2025) & \code{music therapy}; \code{premature infant}; \code{standard care}; \code{physiological parameters}; \code{weight gain}; \code{white matter}; \code{Bayley scales} \\
\bottomrule
\end{tabularx}
\end{table}

Across all periods, \code{premature infant} and \code{music therapy} remained consistently prominent, indicating a continuing focus on preterm populations and music-based care. The temporal pattern nevertheless suggests a shift in emphasis: earlier studies were dominated by physiological regulation and stress-related indicators, whereas later periods incorporated more intervention-specific, developmental, and longer-term outcome language. In particular, recent periods showed greater visibility of terms related to parental involvement, standard-care comparison, and neurodevelopmental follow-up. Taken together, these findings suggest that NICU music intervention research has evolved from an early concentration on physiological and behavioral outcomes toward a broader framework that includes intervention delivery, developmental care, and longer-term outcomes.

\LiteralHeading{Keywords by intervention type}

A RAKE-based keyword analysis was conducted to compare differences in emphasized concepts according to intervention type. The results showed that although MM and MT studies shared some core keywords, they differed in their primary research focus.

\begin{table}[ht]
\centering
\caption{Representative RAKE-derived keyphrases by intervention type}
\label{tab:mmmtkeywords}
\begin{tabularx}{\textwidth}{>{\raggedright\arraybackslash}X >{\raggedright\arraybackslash}X}
\toprule
Music medicine (MM) & Music therapy (MT) \\
\midrule
\code{premature infant}; \code{heart rate}; \code{respiratory rate}; \code{physiological parameters}; \code{weight gain}; \code{classical music}; \code{mother lullaby}; \code{mother voice} & \code{music therapy}; \code{premature infant}; \code{infant bonding}; \code{mother anxiety}; \code{behavioral state}; \code{Bayley scales}; \code{depressive symptoms}; \code{hospital discharge} \\
\bottomrule
\end{tabularx}
\end{table}

The contrast between intervention types is summarized in Table 4. MM-related studies were more strongly associated with physiological regulation, growth, and passive auditory stimulation, whereas MT-related studies showed greater prominence of relational, developmental, and psychosocial keyphrases. Although both groups shared core terms related to preterm infants and physiologic outcomes, MT displayed a broader semantic range linked to parent-infant interaction and longer-term developmental follow-up.

\SubHeading{Topic modeling}

\LiteralHeading{Topic modeling results}

The optimal number of topics was determined using the \code{CaoJuan2009}, \code{Arun2010}, and \code{Deveaud2014} metrics. In general, lower values for \code{CaoJuan2009} and \code{Arun2010} and higher values for \code{Deveaud2014} indicate better topic separation. Although models with fewer topics showed partial improvement in some metrics, they produced overly simplified structures that merged developmental, physiological, and relational content. By contrast, models with more topics yielded smaller and less stable themes that were difficult to interpret consistently across periods and intervention types. A four-topic solution therefore provided the most parsimonious and clinically interpretable representation of the corpus.

\begin{table}[ht]
\centering
\caption{Four-topic solution and representative topic terms}
\label{tab:topicmodel}
\begin{tabularx}{\textwidth}{c >{\raggedright\arraybackslash}X >{\raggedright\arraybackslash}X}
\toprule
Topic & Representative terms & Interpretation \\
\midrule
1 & \code{kangaroo care}; \code{mother anxiety}; \code{behavioral state}; \code{physiological responses}; \code{autonomic nervous system}; \code{stress level} & Parent-involved physiological regulation and stress reduction \\
2 & \code{weight gain}; \code{gestational age}; \code{salivary cortisol}; \code{mother voice}; \code{music intervention}; \code{bonding scale} & Growth and developmental outcomes of music intervention \\
3 & \code{respiratory rate}; \code{physiological parameters}; \code{classical music}; \code{mother lullaby}; \code{vital sign}; \code{birth weight} & Immediate physiological responses to music stimuli \\
4 & \code{music therapy}; \code{infant bonding}; \code{depressive symptoms}; \code{toddler development}; \code{hospital discharge}; \code{Bayley scales} & Developmental and psychosocial effects of music therapy \\
\bottomrule
\end{tabularx}
\end{table}

The four-topic solution and representative terms are summarized in Table 5. Taken together, the topic structure suggests that NICU music-intervention research is organized around partially overlapping domains of physiological stabilization, parent-infant co-regulation, broader intervention effects, and developmental or psychosocial follow-up. The highest-probability terms for each topic, ranked by beta values, are presented visually in Figure 2.

\begin{figure}[ht]
\centering
\includegraphics[width=0.9\textwidth]{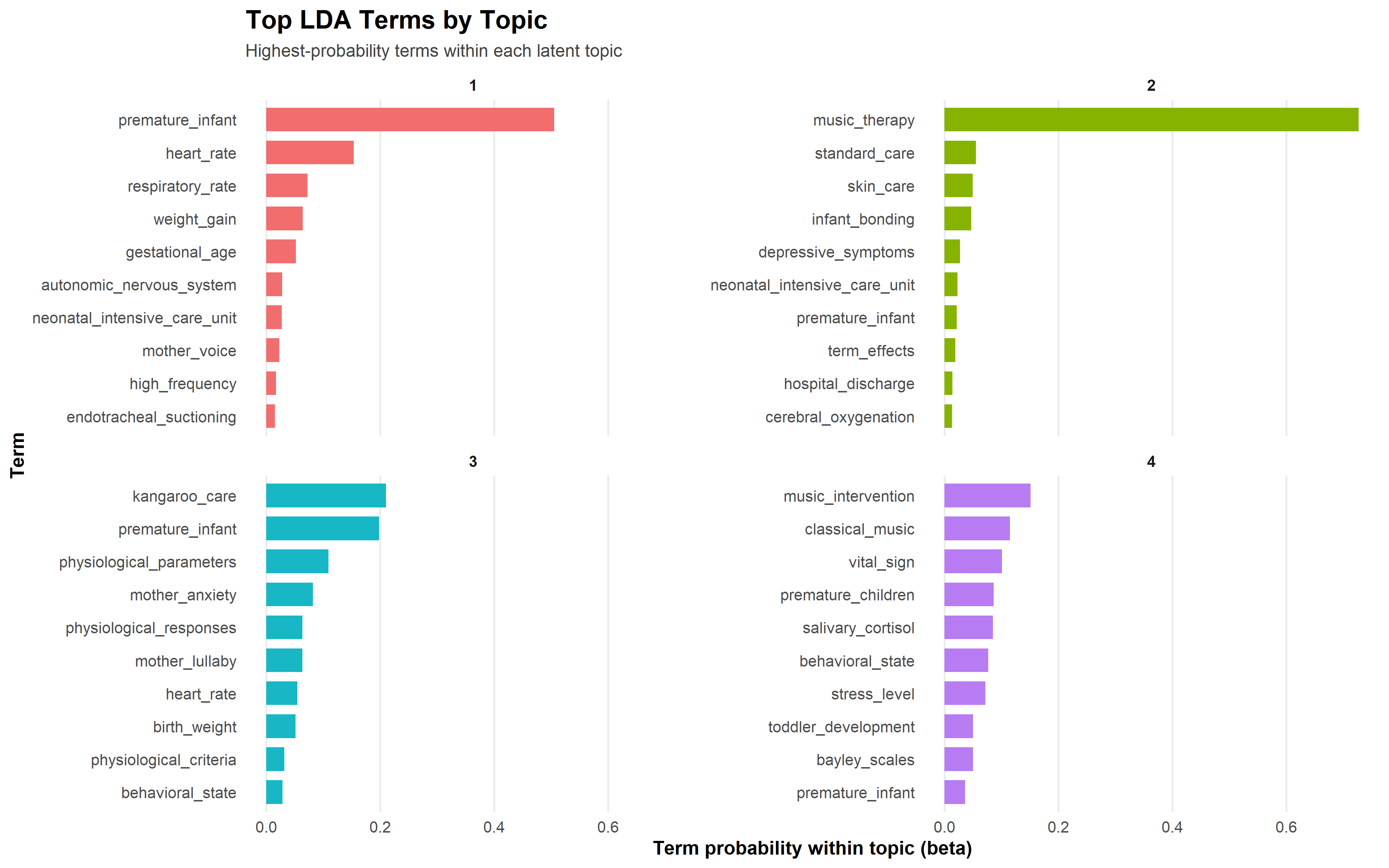}
\caption{Top LDA terms for each topic. Bars represent the highest-probability terms within each latent topic, based on beta values, and illustrate the semantic structure used to interpret the four-topic solution.}
\end{figure}

\LiteralHeading{Topic trends}

To examine changes in research topics over time, publications were divided into four publication periods based on year so that each period contained a roughly comparable number of studies, and each study was assigned to the topic with the highest gamma probability. The number of studies assigned to each dominant topic was then compared across periods (Figure 3). The results revealed clear temporal shifts in topic distribution.

Topic 1 accounted for the largest proportion of studies during the early periods, indicating that parent-infant interaction, physiological stability, and stress responses were major concerns in earlier NICU music intervention research. Topic 2 showed a gradual increase over time and peaked in Period 3, suggesting growing interest in developmental and growth-related outcomes. Topic 3 showed a relatively lower overall proportion with fluctuations across periods. Topic 4 maintained a comparatively low proportion throughout the entire period but exhibited a gradual upward trend, especially in Periods 3 and 4, indicating that developmental and psychosocial effects of music therapy have become an emerging area of interest in recent years.

Taken together, these findings indicate that NICU music intervention research initially concentrated on a limited number of core themes but has become increasingly diversified over time. Additional visualizations of topic similarity, keyword connectivity, and topic distribution by intervention type and publication period are provided in Additional file 1: Figures S1--S4. Supplementary summaries of effect-focused phrase extraction are provided in Additional file 1: Figure S5 and Tables S2--S3.

\begin{figure}[ht]
\centering
\includegraphics[width=0.9\textwidth]{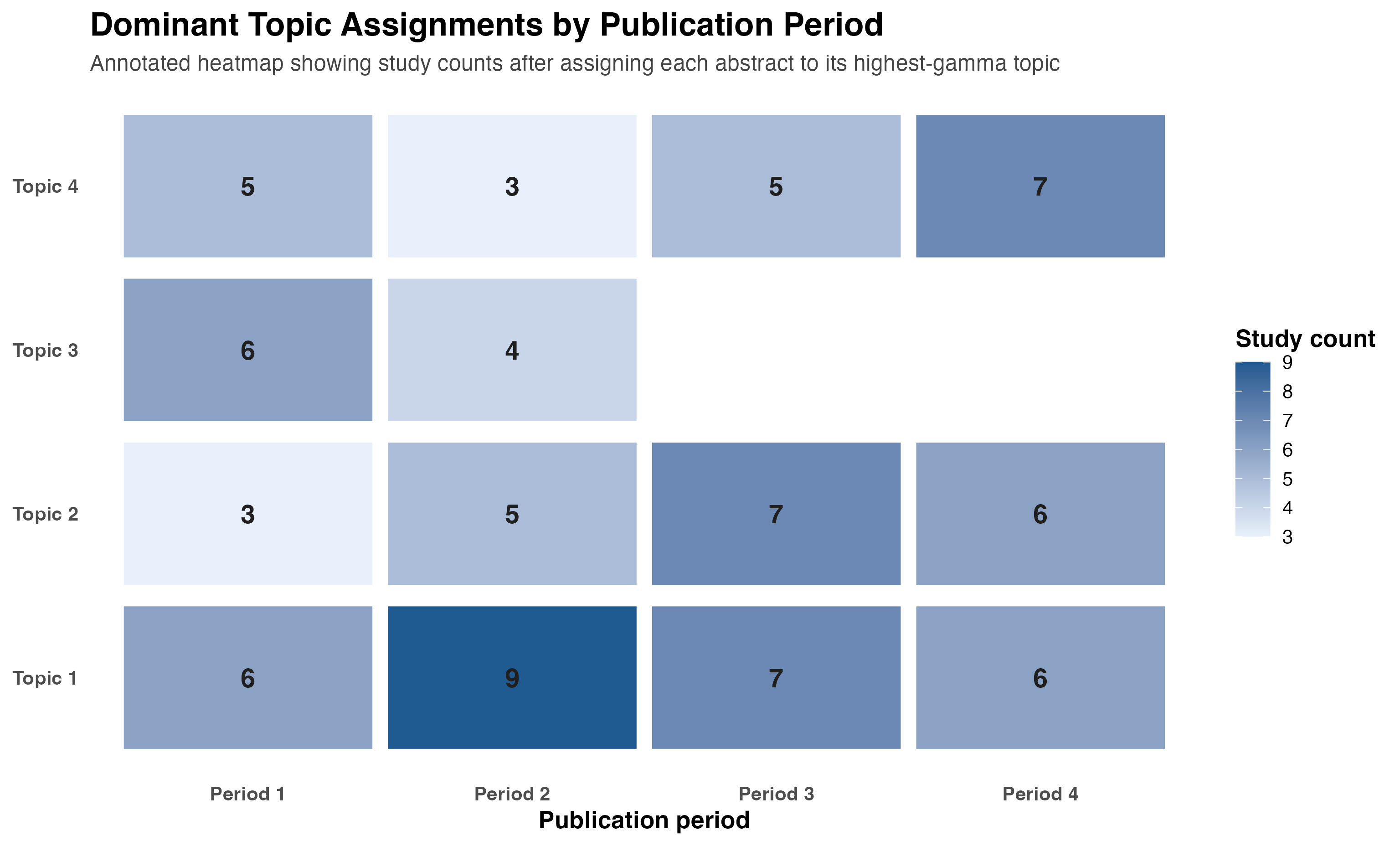}
\caption{Dominant topic assignments by publication period. Each study was assigned to the topic with the highest gamma probability, and the annotated heatmap shows how many studies were assigned to each dominant topic in each period.}
\end{figure}

\MainHeading{Discussion}

This study applied text mining and topic modeling to 83 abstracts of NICU music-based intervention studies published between 1998 and 2025 in order to examine research trends and thematic structures in the field. The findings showed not only a steady increase in the volume of research, but also increasing diversification in both research topics and intervention approaches. These results suggest that music in NICU settings is expanding beyond a simple supportive stimulus toward an intervention encompassing developmental, relational, and psychosocial goals.

First, the analysis by publication period showed that more studies were accumulated within shorter time spans in recent years, indicating rapidly increasing interest in NICU music-based interventions. Early studies mainly focused on immediate physiological stability in preterm infants, such as heart rate, respiratory rate, and stress-related responses. More recent studies, however, have increasingly addressed broader issues, including neurodevelopment, long-term developmental outcomes, parental emotional well-being, and parent-infant interaction. This pattern is consistent with current clinical trends that place growing emphasis on non-pharmacological interventions and family-centered care in NICU practice.

Second, the comparison between MT and MM revealed both shared and distinct characteristics. MM studies were mainly centered on relatively passive auditory stimulation, such as music listening, lullabies, and the mother's voice, and tended to focus on physiological stability-related outcomes, including heart rate, respiratory rate, vital signs, weight gain, and stress-related indicators. In contrast, MT studies encompassed broader domains, including bonding, anxiety, emotional support, family well-being, and developmental outcomes, alongside music therapy, singing, and interactive musical experiences. These findings support the view that MT involves not only the provision of musical stimuli but also relational and therapeutic processes. At the same time, the continued overlap in the use of MT and MM terminology suggests the need for clearer conceptual definitions and reporting standards in future research.

Third, topic modeling showed that NICU music intervention research can be structured into four major themes. Topic 1 reflected parent-involved physiological regulation and stress reduction, Topic 2 represented growth and developmental outcomes, Topic 3 captured immediate physiological responses to music stimuli, and Topic 4 reflected the developmental and psychosocial effects of music therapy. The relatively high similarity between development- and physiology-related topics suggests that physiological stability and developmental outcomes are closely connected in NICU music intervention research. In contrast, the psychosocial music therapy topic appears comparatively independent, indicating that music therapy-based research has developed as a distinct area emphasizing parent-infant relationships, bonding, long-term development, and psychosocial support.

Fourth, when the changes in intervention types and topic distributions across periods are considered together, NICU music intervention research appears to have shifted from early exploratory and passive listening-based approaches toward more active and interactive forms of intervention, including music therapy, live music, singing, and parent-involved interventions. This trend suggests that contemporary NICU music-based interventions are increasingly designed not only for environmental modulation or soothing, but also to address infant development and parental emotional experience in a more integrated manner.

These findings have several clinical implications. First, intervention type and provider role should be more clearly distinguished when music-based interventions are implemented in NICU settings. Second, the role of credentialed music therapists may be particularly important when intervention goals include parent participation, relationship building, and psychosocial support. Third, future NICU music intervention programs may be most effective when embedded within interdisciplinary care models involving nurses, physicians, therapists, and families.

This study is meaningful in that it provides a systematic overview of NICU music intervention literature from a text mining perspective, but several limitations should be considered. First, the corpus comprised 83 abstracts, and the analysis was therefore intended to describe conceptual patterns and thematic shifts in the available literature rather than to provide a definitive representation of the entire field. At the same time, this relatively modest corpus reflected a deliberately focused set of peer-reviewed NICU music-intervention studies, which helped preserve topical specificity and clinical relevance in the trend analysis. In addition, because the review was limited to English- and Korean-language publications, the findings may not fully represent trends reported in other linguistic or regional research contexts. Although the analysis relied on abstracts rather than full texts, the use of normalized RAKE-based keyphrases helped preserve clinically meaningful multi-word concepts and reduce fragmentation of related expressions, thereby strengthening conceptual interpretation. However, abstract-level analysis may still have limited the capture of detailed intervention protocols and contextual variation across studies. Second, preprocessing and keyword refinement involved some degree of researcher judgment. Third, because topic modeling is probabilistic, the resulting topic structure may vary depending on model settings. Future studies should incorporate full-text analyses and more refined comparisons based on intervention type, provider qualification, and clinical context.

\MainHeading{Conclusions}

This study analyzed 83 NICU music-based intervention studies published between 1998 and 2025 using text mining and topic modeling to identify research trends and thematic structures. The findings showed that research in this field has increased rapidly in recent years and that its focus has expanded from early physiological stabilization toward developmental, relational, and psychosocial domains. In addition, music medicine primarily focused on passive auditory stimulation and immediate physiological outcomes, whereas music therapy addressed a broader range of topics, including parent-infant interaction, long-term development, and emotional support. Topic modeling identified four major themes: parent-involved physiological regulation, growth and developmental outcomes, immediate physiological responses, and the developmental and psychosocial effects of music therapy.

Overall, NICU music-based intervention research is evolving in an increasingly multidimensional and interdisciplinary direction. Future research and clinical application should aim to clarify the conceptual distinction between music therapy and music medicine, establish more standardized reporting systems for intervention types and outcome domains, and strengthen the role of credentialed music therapists and interdisciplinary collaboration. Such efforts may contribute to more effective support for the complex needs of infants and families in NICU settings.

\MainHeading{List of abbreviations}

\begin{tabular}{@{}ll@{}}
LDA: & Latent Dirichlet allocation \\
MM: & Music medicine \\
MT: & Music therapy \\
NICU: & Neonatal intensive care unit \\
RAKE: & Rapid Automatic Keyword Extraction \\
\end{tabular}

\MainHeading{Declarations}

\SubHeading{Ethics approval and consent to participate}

Not applicable. This study analyzed published literature and did not involve human participants or identifiable personal data.

\SubHeading{Consent for publication}

Not applicable.

\SubHeading{Availability of data and materials}

The datasets analyzed during the current study were derived from published abstracts identified through database searches. Additional preprocessing details, supplementary figures, and summary tables are provided in Additional file 1, included with this arXiv submission as ancillary material. The R code and processed summary data used for figure generation can be accessed at \url{https://github.com/mijeong-kim/nicu-music-textmining}. Further analytic details are available from the corresponding author on reasonable request.

\SubHeading{Competing interests}

The authors declare that they have no competing interests.

\SubHeading{Funding}

This research received no specific grant from any funding agency in the public, commercial, or not-for-profit sectors.

\SubHeading{Authors' contributions}

SK conceived the study and developed the research idea. MK conducted the R-based data analysis. MC prepared the dataset and study materials and drafted the manuscript. All authors reviewed, revised, and approved the final manuscript.

\SubHeading{Acknowledgements}

Not applicable.

\MainHeading{Supplementary material}

Additional file 1 (PDF supplementary methods, figures, and tables for preprocessing, topic modeling, and effect-focused phrase analysis) is included with this arXiv submission as ancillary material.

\end{document}